\documentstyle[prl,aps,epsf,multicol]{revtex}
\newcommand{\be}{\begin{equation}}
\newcommand{\ee}{\end{equation}}
\newcommand{\bea}{\begin{eqnarray}}
\newcommand{\eea}{\end{eqnarray}}
\begin{document}

\title{Traveling Front Solutions to Directed Diffusion Limited Aggregation, Digital Search Trees
and the Lempel-Ziv Data Compression Algorithm}
\author{Satya N. Majumdar}
\address{ 
Laboratoire de Physique Th\'eorique (FER 2603 du CNRS),
Universit\'e Paul Sabatier, 31062 Toulouse Cedex, France }
\maketitle
\widetext

\begin{abstract} 

We use the traveling front approach to derive exact asymptotic results for the
statistics of the number of particles in a class of directed diffusion limited
aggregation models on a Cayley tree. 
We point out that some aspects of these models
are closely connected to two different problems in computer science, namely
the digital search tree problem in data structures and the Lempel-Ziv algorithm
for data compression. The statistics of the number of particles studied here is related
to the statistics of height in digital search trees which, in turn, is related to the statistics 
of the length of the longest word formed by the Lempel-Ziv algorithm. Implications of our results
to these computer science problems are pointed out.

\end{abstract}
\begin{multicols}{2}

\section{Introduction}

The simple model of diffusion limited aggregation (DLA), ever since its introduction by Witten 
and Sander in 1981\cite{WS}, has continued to play a central role in understanding the fractal
growth phenomena. Besides raising a number of conceptual issues regarding pattern formation, this
model has also found numerous applications in physical processes ranging from dielectric 
breakdown\cite{DB} and Hele-Shaw fluid flow\cite{HS} to electrodeposition\cite{Electro}
and dendritic growth\cite{Vicsek}. In the simplest version of this model, one considers, for 
example, a square lattice where the origin is a seed. Particles are released sequentially from 
the boundary. Each particle performs a random walk in space and when it comes in contact with
the growing cluster around the central seed, it sticks to the cluster and thus the cluster
grows. This growing DLA cluster has a fractal structure with many branches that are separated by 
deep `fjords'. Despite various advances, characterizing this fractal pattern 
quantitatively has 
remained a major theoretical
challenge for the past two decades\cite{Halsey}. One clear picture that has emerged out of 
various studies is that the key effect responsible for this complex pattern is the dynamical
`screening' : a newly arriving particle has more probability to attach to the `tip' sites
compared to other boundary sites that are deep inside a `fjord'. As a result, the faster 
growing
parts of the cluster boundary shield or screen other boundary sites from further growth.    

To understand this dynamical screening effect more quantitatively, it is desirable to construct 
a simpler model which incorpoartes the screening effect and yet is analytically tractable. 
Bradley and Strenski\cite{BS} introduced such a model where particles undergo a directed
diffusion limited aggregation (DDLA) on a Cayley tree. Physically this corresponds to the
situation when there is a strong external field such as the gravity or an electric field
which forces the particles to choose an overall direction of motion.
In this 
DDLA model, one starts with
a Cayley tree of height $l$ (see Fig. 1) where all the $2^{l}-1$ sites of the lattice
are initially empty. Then a particle is introduced at the top site and it performs
a directed (downwards) random walk (from any site it choooses one of the two daughter
sites at random and moves there) till it reaches one of the bottom leaves and can descend no 
more. It then occupies that leaf site. 
\begin{figure}
  \narrowtext\centerline{\epsfxsize\columnwidth \epsfbox{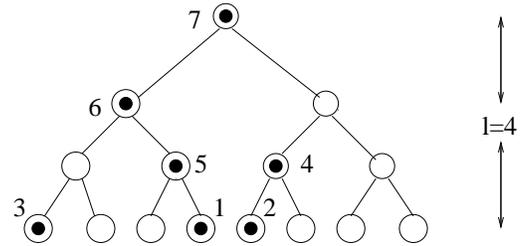}}
\caption{A typical history of the DDLA process till saturation on a Cayley tree with height 
$l=4$. The occupied 
nodes 
contain a black filled circle inside them and the number next to an occupied site
denotes the particle number.} 
\end{figure}
\noindent
Then a
second particle is introduced from the top site and it also performs a directed
random walk. This second particle will stop if it happens to reach a 
site such that atleast one of the daughter nodes of that site is already occupied. It can not
descend any more and it rests at that site for all subsequent times. Note that in this model, 
each site can contain 
at most one particle. Then a third particle is released from the top and so on. Basically,
after reaching any site, say the $i$-th site, a particle attempts to hop down to one of the two 
daughter 
nodes of $i$ and it actually moves to the target site provided both the daughter nodes of 
$i$ are empty. If atleast one 
of them is occupied, the particle can not descend any further and it rests at site $i$
forever.  Then the next particle is added and the process continues 
until no more particles can be put in, i.e. when the top site gets occupied. The tree is then
said to be saturated. One such history of the process till its saturation is shown in Fig. 1.
A typical snapshot of the saturated tree (see Fig. 1) shows that the cluster has voids
of various sizes. A useful quantity to characterize the pattern of the cluster is the
total number of particles $n_l$ in the saturated tree. Clearly $n_l$ is a random variable,
varying from one history of the process to another. The quantities of central importance 
in this problem are the average density,
$\rho_l= {\langle n_l\rangle}/[2^{l}-1]$ and the fluctuations of $n_l$ around
its average value. How do these two quantities behave asymptotically for a large tree, i.e.
when $l\to \infty$ ?

While it was easy to write down the basic recursion relations (see later in Section II) for
certain probabilities associated with the DDLA process on a tree,
they turned out to be nonlinear\cite{BS} and hence it was difficult
to determine even the asymptotic behavior of $\rho_l$ for large $l$. Bradley and Strenski studied
the recursion relations numerically and found, somewhat unexpectedly, that $\rho_l$ decays slower 
than exponentially with $l$ for large $l$ but the precise nature of this decay was not evident 
from 
their numerics\cite{BS}. Later Aldous and Shields\cite{AS} studied via rigorous probabilistic 
methods a 
completely different model namely a continuous time version of the so called digital search tree
(DST) problem relevant in computer science\cite{Knuth,Mahmoud,AS,FS,FR,KPS,LS}. As we will see 
later in 
Section-VI, these 
two models namely the DDLA and the DST share the same recursion relation, though for very 
different quantities. The rigorous results of Aldous and Shields\cite{AS}, when translated back 
in the DDLA language, would indicate a streched exponential decay for the average density,
$\rho_l \sim 2^{-\sqrt{2l}}$ for large $l$. Recently a more refined result on DST was derived
by Knessl and Szpankowsky\cite{KS}. Unaware of the DDLA or the DST model, Hastings
and Halsey also studied independently a related model recently\cite{HH} and used extremal 
arguments to conclude the same streched exponential decay for the average density.  

The methods used by the mathematicians, though rigorous, lack physical transparency. On the
other hand, the extremal arguments used by Hastings and Halsey, though physically intuituive,
are heuristic. Moreover, it is not easy to derive quantitative results regarding fluctuations
of the number of particles $n_l$ via these methods. For example, how does the width
$w(l)=\sqrt{\langle n_l^2\rangle-{\langle n_l\rangle}^2}$ behave as
a function of $l$?  
Besides, none of these methods are easily adaptable to extract the
leading asymptotic behaviors in more general models such as one that we will consider
in this paper. Our approach in this paper would be to use 
the powerful techniques (suitably adapted for our problem) 
of traveling fronts, originally developed in the context of nonlinear reaction-diffusion
systems and population dynamics\cite{KPP,Fisher,Murray}. 
The techniques of traveling fronts have found a
host of very useful applications\cite{UVS}. Recently we have pointed out that in many
extreme value problems in both physics and computer science, one can successfully use
the traveling front techniques to derive exact asymptotic results for the statistics
of the extremum of a set of correlated random variables\cite{TF,bst1,bst2,frag}. The
present paper points out yet another useful application of the traveling 
front techniques, namely in a generalized DDLA problem with relevance to a 
class of search tree problems in computer science.

The traveling
front method, though technically not rigorous in the strict mathematical sense, has
the following advantages over the other methods used in the DDLA problem: (i) this method
is not model specific, is more general and is easily adapdable to more general models
such as the ones that will be studied in this paper, (ii) it is easy to implement and is 
physically
transparent and (iii) it provides a very cheap way to extract the leading asympotic behavior
exactly without using too much mathematics. Besides rederiving the known results in
the standard DDLA problem, this method also allows us to derive many new results
in more generalized models. For example, we show that in the DDLA model, the 
random variable $n_l$ approaches to its average value, $n\to \langle n_l\rangle$ in
the $l\to \infty$ limit. This is the example of the extreme concentration of measure, i.e.,
the distribution of $n_l$ tends to a delta function. In particular, we show that the
width of the distribution decays slowly as a power law, $w(l)\sim l^{-1/2}$ as $l\to \infty$.
Furthermore, we point out the close connection between the DDLA problem and  
the DST problem widely studied by computer scientists\cite{Knuth,Mahmoud,AS}. The
later problem is also related to the well known Lempel-Ziv algorithm used in data 
compression\cite{LZ}.
Some of the results derived in this paper will constitute
new results in these computer science problems.

The rest of the paper is organized as follows. In the next section we introduce a generalized
b-DDLA model (where $b$ is a positive integer). The original DDLA model of Bradley and Strenski
is a special case of this b-DDLA model with $b=1$. We then derive the asymptotic statistics
of the number of particles in the b-DDLA model using the traveling front technique, suitably 
adapted for this model. In Section III, we generalize these results to the case when the 
Cayley tree has $m>2$ branches. Section IV considers the DDLA model with a bias when
a particle has more probability to go to the left branch compared to the right one. 
In Section V, we point out the detailed connections between
the DDLA model, the DST problem in computer science, and the Lempel-Ziv parsing algorithm
and discuss the implications of our results for the generalized b-DDLA model in the
context of computer science. Finally a brief summary and a conclusion along with
a list of open problems are offered in Section VI. 

\section{The b-DDLA Model and Its Traveling Front Solution}

Here we introduce a generalized b-DDLA model where the `hard' screening of the usual
DDLA model is `softened' in the following sense. As in the usual DDLA model, 
one starts with an empty Cayley
tree of height $l$ and the particles are introduced sequentially at the top site
and they go down the tree one at a time by performing a random walk. 
However, now each site can contain at most $b$ particles where $b$ is a positive integer. During 
its journey downward, when a new 
particle arrives
at an empty site, say the $i$-th site, it tries to move to one of the daughter nodes of $i$
chosen at random. It actually moves to the target site provided both the daughter sites
contain less than $b$ particles. If either of them contains $b$ particles, i.e. 
is completely full, then the incoming particle can not move down any further and it then stays
at site $i$ forever. Thus, in this model, a site can incorporate `screening' if and only if 
it has full capacity, i.e. when it has $b$ particles. Otherwise it fails to screen. This model
thus mimics the physical situation when one single particle is incapable of stopping an incoming
particle to go down, but the screening comes into play only as a collective effect when there
are $b$ particles in the site. This is like a tunnelling effect, where a particle can go through 
a barrier provided the barrier is not too high. However, the rate of tunnelling goes to zero
when the barrier height crosses a threshold. In our model, the parameter $b$ acts like the 
threshold value. Clearly, for $b=1$ this model reduces to the original DDLA model studied by 
Bradley and Strenski\cite{BS}. 

As in the $b=1$ case, the tree is going to be saturated after a finite number of particles have 
been added to it. This happens when the top site contains $b$ particles. No further particles 
can then be put in. The number of particles $n_l$ required to saturate the tree of height $l$ is 
clearly a random variable, fluctuating from one history of the process to another. The main 
question we address is: what is the statistics of $n_l$ for large $l$? 
In particular, we would compute the average density at saturation, $\rho_l= \langle 
n_l\rangle/[2^{l}-1]$ and the width of
the distribution, $w(l)=\sqrt{\langle n_l^2\rangle-{\langle n_l\rangle}^2}$ for large $l$.

Following Bradley and Strenski\cite{BS} for the $b=1$ case, we define $G_l(n)$ to be the 
probability that the tree of height $l$ is {\em not} saturated after the addition of
$n$ particles, i.e. the top site is not yet filled by $b$ particles after $n$ particles
have been added to the tree. It is easy to see that $G_l(n)$ satisfies the following
recursion relation,
\begin{equation}
G_{l+1}(n+b)={1\over {2^n}}\sum_{n_1=0}^n {n\choose n_1} G_l(n_1)G_l(n-n_1),
\label{recur1}
\end{equation}
for all $l\geq 1$ and $n\geq 0$. It is useful to think of $l$ as `space' and $n$ as `time'.
The Eq. (\ref{recur1}) is supplemented with the `boundary' condition,
$G_1(n)=1$ for all $0\leq n \leq (b-1)$ and $G_1(n)=0$ for all $n\geq b$ and the `initial'
condition, $G_l(n)=1$ for $0\leq n\leq (b-1)$ for all $l\geq 1$. The recursion relation
in Eq. (\ref{recur1}) is easy to understand. Suppose we have added $(n+b)$ particles to a
tree of height $(l+1)$ (the left hand side of Eq. (\ref{recur1})). The condition that the top 
site is not yet filled by $b$ particles indicates
that before the addition of the last $b$ particles, the two daughter nodes of
the top site must have both remained unsaturated. This is becuase, if either one or both of them
had been saturated after the addition of $n$ particles, then any further added particle would 
not be able to go down and would rest 
at the top site, and hence the top site would then get saturated after 
the addition of $(n+b)$ particles. The two daughter nodes are the roots of two
uncorrelated subtrees, each of height $l$. Hence the probability that both
remain unsaturated is given by their product. Also, the number of particles $n_1$
that enter, for example, to the left subtree (out of a total number of $n$ particles 
that enter both subtrees) must have a binomial distribution, thus explaining the right hand side
of Eq. (\ref{recur1}).

Note that for fixed $n>0$, the probability $G_l(n)\to 0$ as $l\to 1$ and $G_l(n)\to 1$
as $l\to \infty$. For later analysis, it turns out to be convenient to define the
complementary probability, $F_l(n)=1-G_l(n)$, that has the opposite behavior as a
function of $l$, namely $F_l(n)\to 1$ as $l\to 1$ and $F_l(n)\to 0$ as $l\to \infty$
for any fixed $n>0$. The quantity $F_l(n)$ denotes the probability that the tree
of height $l$ gets saturated before $n$ particles are added. From Eq. (\ref{recur1}), one finds
that $F_l(n)$ satisfies the recursion,
\begin{eqnarray}
F_{l+1}(n+b)&=&{1\over {2^{n-1}}}\sum_{n_1=0}^n {n\choose n_1} F_l(n_1) \nonumber \\
&-&{1\over 
{2^n}}\sum_{n_1=0}^n {n\choose n_1}F_l(n_1)F_l(n-n_1),
\label{recur2}
\end{eqnarray}
with the boundary condition, $F_1(n)=0$ for $0\leq n\leq (b-1)$ and $F_1(n)=1$ for $n\geq b$
and the initial condition, $F_l(n)=0$ for $0\leq n\leq (b-1)$ for all $l\geq 1$. It is useful to 
think by fixing the `time' $n$ while varying the `space' $l$. Clearly $F_l(n)\to 0$ as $l\to 
\infty$, since almost surely a tree of infinite height will not be saturated before the addition
a fixed, finite number $n$ of particles. On the other side, for fixed $n$, $F_l(n)\to 1$
as $l\to 0$. For a given fixed $n$, as one increases $l$ from $0$ to $\infty$, the function
$F_l(n)$ starts off at the value $1$ at $l=0$ and then drops off to $0$ beyond some
characteristic length scale $l^*(n)$, as shown schematically in Fig. (2). As $n$ increases, this 
characteristic length scale $l^*(n)$ also increases (see Fig. (2)), thus giving rise to
a traveling front structure with the front located at $l^*(n)$. In fact, we will see later that 
in the limit of large $l$
(when one can treat $l$ as a contiuous variable) and large $n$, the width $w(n)$ of the
front goes to zero, indicating that asymptotically the function $F_l(n)$ becomes
a Heaviside theta function, $F_l(n)\to \theta(l^*(n)-l)$.  
\begin{figure}
  \narrowtext\centerline{\epsfxsize\columnwidth \epsfbox{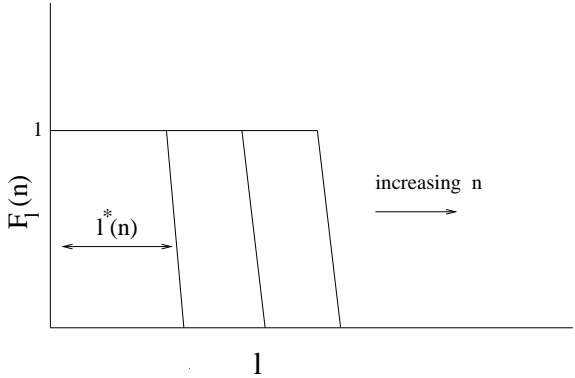}}
\caption{Schematic behavior of the probability $F_l(n)$ as a function of $l$ for different
fixed values of $n$. The three curves correspond to different values of $n$ increasing
from left to right. The $y$ axis is dimensionless while the $x$ axis has arbitrary units.}
\end{figure}
\noindent

Note that Eq. (\ref{recur2}) is nonlinear and hence is diffcult to solve exactly.
However, the exact asymptotic informations regarding the position $l^*(n)$ of the front
and its width can be deduced by adapting the  
traveling front techniques that were originally devised to deal with nonlinear partial 
differential differential equations in reaction-difffusion systems\cite{KPP} and pupulation 
dynamics\cite{Fisher,Murray}. The basic idea behind this approach is very simple.
If there is a front $l^*(n)$, then ahead of the front $l> l^*(n)$, $F_l(n)$ is very small
and hence one can neglect the nonlinear term (the second term) on the right hand side
of Eq. (\ref{recur2}) and one simply gets a linear recursion,
\begin{equation}
F_{l+1}(n+b)\approx {1\over {2^{n-1}}}\sum_{n_1=0}^{n} {n\choose n_1} F_l(n_1).  
\label{recur3}
\end{equation}
Suppose one could solve this linear equation exactly satisfying the required initial
condition. Now one expects that the solution of the linear equation (\ref{recur3})
and the `real' solution of the nonlinear equation (\ref{recur2}) will coincide
in the regime beyond the front, i,e. for $l>l^*(n)$. On the other hand, the two
solutions will start differing from each other as one decreases $l$ below $l^*(n)$, where
the solution of the nonlinear equation (\ref{recur2}) will tend to $1$ where as
the solution of the linear equation (\ref{recur3}) will grow beyond $1$ with decreasing $l$
(as there is no nonlinear term to control the solution). Thus, as one decreases
$l$ from infinity, the front 
position
$l^*(n)$ is approximately the value of $l$ at which the solution of the linear equation 
becomes $\sim O(1)$. Thus, according to this approach, one first solves the linear equation
(\ref{recur3}) to obtain $F_l(n)|_{\rm linear}$ and then reads off the front position $l^*(n)$ 
from the condition, $F_{l^*(n)}(n)|_{\rm linear}\approx O(1)$. By $O(1)$, one means that at
$l=l^*(n)$, the solution $F_l(n)$ should not diverge or decay exponentially with increasing $n$.
Note that this is a slightly generalized version of the usual traveling front 
method\cite{UVS,TF} where one usually has a linear operator with constant coefficients
which admits an exponentially decaying solution of the form $\exp[-\lambda(x-vt)]$ with constant 
width. The present approach
is more general and works even when the linear operator does not admit an exponentially
decaying solution with a constant width.

Under this traveling front approach , our task thus reduces to solving the linear equation
(\ref{recur3}) which, however, is still nontrivial. To proceed, we define a somewhat unusual
generating function,
\begin{equation}
{\tilde F}_l(s)= \sum_{n=0}^{\infty} F_l(n) {1\over {(1+s)^{n+1}}}.
\label{gf1}
\end{equation}
Using Eq. (\ref{recur3}) one can then show that ${\tilde F}_l(s)$ satisfies
a rather simple recursion in $l$,
\begin{equation}
{\tilde F}_{l+1}(s)= {4\over {(1+s)^b}}{\tilde F}_l(2s),
\label{fls1}
\end{equation}
for all $l\geq 1$. The steps leading to Eq. (\ref{fls1}), starting from Eqs. (\ref{recur3})
and (\ref{gf1}), are not 
completely straightforward.
Hence we present this derivation in the Appendix. The recursion in Eq. (\ref{fls1}) starts
with the initial value ${\tilde F}_1(s)$ which needs to be calculated separately. Noting
that $F_1(n)=0$ for $0\leq n\leq (b-1)$ and $F_1(n)=1$ for $n\geq b$, we find from the 
definition
in Eq. (\ref{gf1}), ${\tilde F}_1(s)= 1/[s(1+s)^b]$. Iterating Eq. (\ref{fls1}) and using the 
expression for $l=1$, we get
\begin{eqnarray}
{\tilde F}_l(s)& =& {2^{l-1}\over {s{\left[(1+s)(1+2s)\dots (1+2^{l-1}s)\right]}^b}}\nonumber \\
&=& {2^{l-1}\over {s}}\exp\left[ -b\sum_{k=0}^{l-1}\ln {(1+2^{k}s)}\right],
\label{fls2}
\end{eqnarray}
for all $l\geq 1$.

We then write the sum inside the exponential in Eq. (\ref{fls2}) in two parts using the
Euler-Maclaurin summation formula, $S(l,s)=\sum_{k=0}^{l-1}\ln {(1+2^ks)}= I(l,s)+R(l,s)$
where $I(l,s)=\int_0^{l}\ln {(1+2^x s)}dx$ and $R(l,s)=S(l,s)-I(l,s)$ is the residual term. 
The integral $I(l,s)$ can be done and one gets, $I(l,s)=[Y(2^ls)-Y(s)]/{\ln 2}$ where the
function $Y(z)$ is given by
\begin{equation}
Y(z)=\cases
{ \sum_{n=1}^{\infty} (-1)^{n-1}z^{n-1} n^{-2} \quad\quad  &${\rm for}\,\, z \leq 1$,\cr
{1\over {2}}\ln^2 z -\ln (1+1/z)+\ln {2} + {\pi}^2/12 \quad &${\rm for}\,\, z >1$. \cr} 
\label{yz}
\end{equation} 
The structure of these expressions suggests a natural scaling limit, $s\to 0$, $l\to \infty$
but keeping the product $z=s2^l$ fixed but arbitrary. We also treat $l$ as a continuous
variable in this limit. Furthermore, we focus only near the tail of the scaling regime, i.e.
when $z>>1$. In this regime, it is sufficient to keep only the first term in the second line
of Eq. (\ref{yz}) for the expression of $Y(z)$. Besides, one can also check that the
residual term is subleading in this regime. Keeping only the leading terms we get  
\begin{equation}
{\tilde F}_l(s)\approx {1\over {s}}
\exp\left[\ln {2}\left( l-{b\over {2}}{\left(l +
{\log}_2\left(s\right)\right)}^2\right)\right].
\label{fls3}
\end{equation}
We still need to invert the generating function in Eq. 
(\ref{gf1}) to obtain the asymptotic behavior of $F_l(n)$. The scaling limit
corresponds to taking $n\to \infty$, $l\to \infty$ but keeping the ratio $2^l/n$ 
fixed but arbitrary. Using Eq. (\ref{fls3}) and inverting Eq. (\ref{gf1}) (using
the Bromwich inversion formula and then using the standard steepest descent method), we find the 
following leading asymptotic behavior,
\begin{equation}
F_l(n)\approx \exp\left[\ln {2}\left( l-{b\over {2}}{\left(l - 
{\log}_2\left(n\right)\right)}^2\right)\right],
\label{fln1}
\end{equation}
valid in the tail $l>> {\log}_2(n)$.

Having obtained the asymptotic solution (\ref{fln1}) of the linear equation (\ref{recur3}),
the location of the front $l^*(n)$ can be read off from the equation, $F_{l^*}(n)|_{\rm 
linear}\sim O(1)$. Using this criterion in Eq. (\ref{fln1}), we find that the front position
$l^*(n)$ is given by the quadratic equation,
\begin{equation}
l^* - {b\over {2}}[l^*- {\log}_2(n)]^2 =0.
\label{l*1}
\end{equation}
As we decrease $l$ from $\infty$, we will encounter the larger root first, which will
correctly locate the front position. From Eq. (\ref{l*1}), we get, for large $n$, the  
asymptotic front position
\begin{equation}
l^*(n) \approx {\log}_2(n)+ \sqrt{{2\over {b}}{\log}_2(n)}.
\label{l*2}
\end{equation}  
Furthermore, substituing $l= l^*(n) +y$ in Eq. (\ref{fln1}) and expanding for small $y$, we find
to leading order, $F_l(n)\sim \exp\left[-\sqrt{2b\ln (2)\ln (n)}\left(l-l^*(n)\right)\right]$,
indicating that the width of the distribution, characterizing the fluctuation of $l$ around its 
average value $l^*(n)$, decreases extremely slowly with $n$ as, 
\begin{equation}
w(n)\approx 1/{\sqrt{2b\ln (2)\ln (n)}},
\label{wn}
\end{equation}
as $n\to \infty$. The fact that the width vanishes in 
the $n\to \infty$ limit shows 
that the probability $F_l(n)\to \theta (l^*(n)-l)$, thus indicating an extreme concentration of 
measure, i.e. the random variable $l\to l^*(n)$.

In the above analysis, we kept $n$ fixed and studied the behavior of
$F_l(n)$ as a function of $l$. Alternately, as is more suited for the DDLA problem,
one can keep $l$ fixed and vary $n$. It follows from Eq. (\ref{fln1}) that in the
scaling limit mentioned above, the random variable ${\log}_2 (n)$ approaches to its mean value
$\langle {\log}_2(n)\rangle= l-\sqrt{2l/b}$. Due to the extreme concentration of measure, it 
follows then that $n\to \langle n_l\rangle \approx 2^{l-\sqrt{2l/b}}$ in the scaling limit.
This means that the average density $\rho_l=\langle n\rangle/(2^{l}-1)$ decays as a stretched 
exponential for large $l$,
\begin{equation}
\rho(l) \approx 2^{-\sqrt{2l/b}}.
\label{rhol}
\end{equation}
Besides, substituting ${\log}_2 (n)= \langle {\log}_2(n)\rangle - y_1$ in Eq. (\ref{fln1})
and expanding for small $y_1$, we find, $F_l(n)\approx \exp\left[-\ln (2) \sqrt{2bl}\, y_1\right]$.
This indicates that as a function of $l$, the width of the random variable ${\log}_2(n)$
around its average value $\langle {\log}_2(n)\rangle= l-\sqrt{2l/b}$ decreases algebraically
for large $l$
\begin{equation}
w(l)\approx {1\over {\sqrt{2b \ln^2 (2) l}}}.  
\label{wl}
\end{equation}
Note that the leading order behaviors of the widths in Eqs. (\ref{wn}) and (\ref{wl}) are 
compatible with each other with the 
identification $n\approx 2^l$. The Eqs. (\ref{l*2}), (\ref{wn}), (\ref{rhol}) and (\ref{wl})
constitute the main results of this section.

\section{Generalization to a Tree with $m$ branches}

In this section we generalize our results for the b-DDLA in the previous section (obtained for a 
tree with
$m=2$ branches) to a tree with $m\geq 2$ branches. In this case, during its downward journey
from the top, a particle from a given site attempts to hop to any of the $m$ daughter nodes
with equal probability $1/m$ and can actually hop to the target site provided none of the
$m$ daughter nodes is full with $b$ particles. If it fails to hop, the particle stays at its
current site for all subsequent time. The probability $G_l(n)$ that a tree of height $l$
is yet to be saturated after the addition of $n$ particles satisfies the generalized recursion 
relation,
\begin{equation}
G_{l+1}(n+b)= {{n!}\over {m^n}}\sum_{{n_i}=0}^{m}\prod_{i=1}^m {G_l(n_i)\over 
{n_i!}},
\label{mrecur1}
\end{equation}
where the variables $n_i$'s satisfy the constraint $\sum_{i=1}^m n_i =m$. Thus the binomial
coefficient in Eq. (\ref{recur1}) of the previous section gets replaced by a multinominal.
The rest of the analysis is straightforward and similar to the previous section. We define as
usual, the complementary probability, $F_l(n)=1-G_l(n)$, which satisfies the recursion,
\begin{eqnarray}
F_{l+1}(n+b)&= &{1\over {m^{n-1}} }\sum_{n_i=0}^m { {n!} 
\over {\prod_{i=1}^m n_i!} } F_l(n_1)\nonumber \\ 
&+& {\rm nonlinear\,\, terms},
\label{mrecur2}
\end{eqnarray}
where we have used the symmetry that all branches are similar to each other.

As before, we solve the equation (\ref{mrecur2}) retaining only the linear terms
and neglecting the nonlinear terms.  We define the generating function as in
Eq. (\ref{gf1}). Following the derivation presented in the Appendix and using the initial
condition, we get the solution
\begin{equation}
{\tilde F}_l(s) ={ {m^{l-1}}\over {s{\left[(1+s)(1+ms)\dots (1+m^{l-1}s)\right]}^b} }.
\label{mfls1}
\end{equation}
The asymptotic analysis is exactly similar to the previous section, except that the proper 
scaling 
limit now is $s\to 0$, $l\to \infty$ but keeping the product $sm^l$ fixed but arbitrary. We 
do not repeat the steps here but present only the final results. We find that as in the $m=2$ 
case, there is
a front whose asymptotic location $l^*(n)$ is given by
\begin{equation}
l^*(n) \approx {\log}_m(n)+ \sqrt{{2\over {b}}{\log}_m(n)},
\label{ml*2}
\end{equation} 
and in the limit $n\to \infty$, the width $w(n)$ of the front vanishes slowly as
\begin{equation}
w(n)\approx 1/{\sqrt{2b\ln (m)\ln (n)}}.
\label{mwn}
\end{equation}
Similarly, we find that for fixed but large $l$, the average density varies as a strectched 
exponential,
\begin{equation}
\rho(l) \approx m^{-\sqrt{2l/b}},
\label{mrhol}
\end{equation}
and the width $w(l)$ of the random variable ${\log}_m(n)$ around its average value
$\langle {\log}_m(n)\rangle=l - \sqrt{2l/b}$ decreases algebraically for large $l$,
\begin{equation}
w(l)\approx {1\over {\sqrt{2b \ln^2 (m) l}}}.
\label{mwl}
\end{equation} 
Note that, interestingly, the asymptotic average value $\langle {\log}_m(n)\rangle=l - 
\sqrt{2l/b}$ is actually independent of $m$.

\section{b-DDLA model with Biased Hopping}

In this section we consider the b-DDLA model on a $m=2$ tree where the particles perform biased random 
walk on their way down the tree. More precisely, when a particle arrives at any 
given site $i$ on its way down 
after being introduced at the top site, it attempts to hop to the left daughter of the node $i$
with probability $p$ and to the right daughter with probability $q=1-p$. As before, it actually moves 
to the target site provided both the daughter nodes have less than $b$ particles. If at least 
one of them is full with $b$ particles, then the particle rests at site $i$ for all 
subsequent times. Then a new particle is added
and the process continues till the top site gets filled with $b$ particles. Once again, we are 
interested in the 
statistics of the number of particles $n_l$ when the tree of height $l$ gets saturated. We define,
as before, $G_l(n)$ to be the probability that the tree of height $l$ remains unsaturated, i.e.
the top site remains unfilled upto the addition of $n$ particles. Following the same logic as in 
Section II, one easily finds the recursion relation,
\begin{equation}
G_{l+1}(n+b)=\sum_{n_1=0}^n {n\choose n_1}p^{n_1}q^{n-n_1} G_l(n_1)G_l(n-n_1),
\label{brecur1}
\end{equation}  
for all $l\geq 1$ and $n\geq 0$. 

The complementary probability, $F_l(n)=1-G_l(n)$, then satisfies
the recursion,
\begin{eqnarray}
F_{l+1}(n+b)&= &\sum_{n_1=0}^n {n\choose n_1} p^{n_1}q^{1-n_1}\left[F_l(n_1)+F_l(n-n_1)\right]\nonumber \\
&+& {\rm nonlinear\,\, terms},
\label{brecur2}
\end{eqnarray} 
with the boundary condition $F_1(n)=0$ for $0\leq n\leq (b-1)$ and $F_1(n)=1$ for $n\geq b$. 
As before, we solve the equation (\ref{brecur2}) keeping only the linear terms and neglecting the 
nonlinear terms. This is done via defining the generating function ${\tilde F}_l(s)$ as in Eq. 
(\ref{gf1}). Following
the same line of derivation presented in the Appendix, we obtain the following recursion 
relation,
\begin{equation}
{\tilde F}_{l+1}(s)= {1\over {p(1+s)^b}}{\tilde F}_l(s/p)+ {1\over {q(1+s)^b}}{\tilde F}_l(s/q),
\label{bfls1}
\end{equation}    
which starts from the initial function, ${\tilde F}_1(s)=1/[s(1+s)]$. One can, in principle, iterate 
Eq. (\ref{bfls1}) starting with $l=1$ and obtain the expressions for ${\tilde F}_l(s)$ for all $l$.
Fortunately, in the scaling regime $s\to 0$, one does not need the full expression for ${\tilde 
F}_l(s)$. Note that in the unbiased case $p=q=1/2$, the appropriate scaling regime was
$s\to 0$, $l\to \infty$ but keeping the product $z=s2^l$ fixed but arbitrary. In the biased case,
it is clear from Eq. (\ref{bfls1}) that the appropriate scaling regime will be set by taking
$s\to 0$, $l\to \infty$ but keeping the product $z=s\sigma^l $ fixed where $\sigma={\rm 
min}(1/p,1/q)$. Thus in this scaling limit, one can approximate Eq. (\ref{bfls1}) by
\begin{equation}
{\tilde F}_{l+1}(s)\approx {{\sigma}\over {(1+s)^b}}{\tilde F}_l(\sigma s).
\label{bfls2}
\end{equation}
The terms neglected in going to Eq. (\ref{bfls2}) from Eq. (\ref{bfls1}) only contribute
to subleading order. Iterating the reduced Eq. (\ref{bfls2}) starting with 
${\tilde F}_1(s)=1/[s(1+s)]$, one obtains
\begin{equation}
{\tilde F}_l(s) \approx {1\over {s\left[(1+s)(1+\sigma s)(1+\sigma^2s)\dots 
(1+\sigma^{l-1}s)\right]^b}}.
\label{bfls3}
\end{equation}

We then invert the transform in Eq. (\ref{bfls3}) using the same asymptotic method as in Section II,
the details of which we do not repeat. The final asymptotic form of the distribution $F_l(n)$
is given by,
\begin{equation}
F_l(n)\approx \exp\left[-{{\ln (\sigma)}\over {2}} 
{\left(l-{\log}_{\sigma}(n)\right)}^2\right],
\label{bfln1}
\end{equation}
valid in the scaling regime, $n\to \infty$, $l\to \infty$ but with the ratio $y=2^l/n$ fixed
at a large value $y>>1$. The front position can be read off from 
the condition $F_{l^*}(n)|_{\rm 
linear}\sim O(1)$ which gives, to leading order, $l^*(n)\approx {\log}_{\sigma}(n)$. Note the 
difference with the unbiased case. Unlike the unbiased case in Section II where the width vanishes 
for large $n$, here the width 
of the distribution remains of $O(1)$ in the large $n\to \infty$ limit, $w(l)\to 1/\sqrt{\ln 
{\sigma}}$. This result also indicates that the average density of particles varies as
$\langle n_l\rangle \approx {\sigma}^l$ for large $l$. Thus, unlike the unbiased case where
the average density decays as a stretched exponential for large $l$, the average density in the biased
case decays exponentially for large $l$,
\begin{equation}
\rho(l)\approx (\sigma/2)^l,
\label{brhol}
\end{equation}
where $\sigma= {\rm min}(1/p,1/q)$. Besides, it follows from Eq. (\ref{bfln1}) that the fluctuations 
of the variable ${\log}_{\sigma}(n)$ 
around its average value $l$ are characterized by a Gaussian tail with width of $O(1)$.    

\section{Connection to Digital Search Trees and The Lempel-Ziv Parsing Algorthim} 

In this section we point out the connection between our generalized b-DDLA model
to the so called digital search tree problem in computer 
science\cite{Knuth,Mahmoud,AS,FS,FR,KPS,Pittel}
which, in turn, is also related\cite{LS} to the Lempel-Ziv data compression algorithm\cite{LZ}.
Suppose we have a data string 
$\{x_1,x_2,\dots, x_n\}$ which needs to be stored 
on a binary tree. According to the digital search tree (DST) algorithm, one proceeds
as follows. Initially all
the nodes of the tree are empty. The first arriving element
$x_1$ is put at the root of the tree. Each node can contain at most one element. 
The second element $x_2$ is put at one of the daughter nodes of
the root chosen at random. Then for the next element $x_3$, one again starts at the root
and choses one of the daughter nodes at random. If the chosen node is empty, $x_3$ goes there.
But if the chosen node, say $i$, happens to be the one that contains $x_2$, then one chooses
one of the two daughter nodes of $i$ at random and puts $x_3$ there. Then one stores the fourth
element $x_4$ following the same algorithm and so on. Essentially each element $x_i$
performs a directed random walk down the tree till it finds an empty site which it then occupies.
The process stops when all the $n$ elements have been stored and the resulting tree is called a 
DST (see Fig. 3). 
Note that according to this DST algorithm, the actual value of a data element say $x_i$ is
not important. This is contrast to other search trees, such as the random binary search 
trees\cite{Knuth,Mahmoud} where the actual value of $x_i$ is used in constructing the tree.
\begin{figure}
  \narrowtext\centerline{\epsfxsize\columnwidth \epsfbox{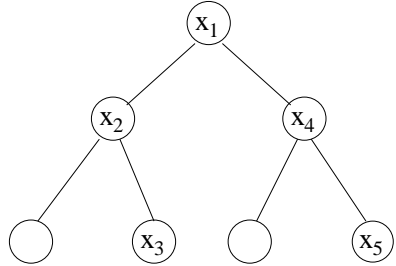}}
\caption{A typical digital search tree constructed from a data string $\{x_1,x_2,x_3,x_4,x_5\}$
of $5$ elements.}
\end{figure}
\noindent

The statistics of various quantities such as the distribution of the 
number of occupied nodes at a given depth (known as the profile of the DST) have been
studied in great detail in the computer science literature\cite{LS}.
Flajolet and Richmond\cite{FR} introduced a generalized version of the DST
where each node can contain at most $b$ elements. In this b-DST, an incoming element $x_i$
performs, as in the $b=1$ case, a directed random walk downwards. However, when $x_i$ 
reaches a new site, say $k$, it will stay forever at $k$ provided the number of already existing 
elements at $k$ is less than $b$. If the site $k$ already contains $b$ elements, then one chooses 
one of the daughter nodes of $k$ and the element $x_i$ hops there. This generalized b-DST
problem has many applications in computer science, notably in the maintenance of
paged hashing tables\cite{FR}. Flajolet and Richmond studied, for example, the average number 
of non-empty nodes in a b-DST as a function of the data size $n$ and the parameter $b$.  

One important characteristic of a b-DST is its height. The height $l$ of a tree with $n$ elements
is defined as the depth, counted from the root, of the farthest element in the tree. Clearly
$l$ is a random variable, fluctuating from one realization of the tree to another and also
it is an extreme variable (denoting the maximum depth). A natural question is: what
is the probability distribution of the height? Let us define $Q_l(n)$ to be the probability
that the height of tree with $n$ elements is $\leq l$. It is easy to see that 
$Q_l(n)$ satisfies the following recursion relation
\begin{equation}
Q_{l+1}(n+b)={1\over {2^n}}\sum_{n_1=0}^n {n\choose n_1} Q_l(n_1)Q_l(n-n_1),
\label{drecur1}
\end{equation}
for all $l\geq 1$ and $n\geq 0$, with the additional condition that $Q_1(n)=1$ for 
all $0\leq n \leq b$ and $Q_1(n)=0$. For the case $b=1$, this recursion relation was
recently studied by Knessl and Szpankowsky\cite{KS}
using rigorous methods. The recurrence in Eq. (\ref{drecur1}) is a 
generalized version of the $b=1$ case and can be understood as follows. Consider
a tree with a total number of $(n+b)$ elements. The first $b$ elements will be stored
in the root and the rest of the $n$ elements will be distributed to the left and right 
daughter subtrees. The probability that one of the subtrees, say the left one, gets
$n_1$ elements out of a total $n$ elements is simply given by the binomial distribution.
Also, since the condition that the height of the full tree is $\leq l+1$ (the left hand side
of Eq. (\ref{drecur1})) indicates that the height of both of the daughter subtrees must be
$\leq l$. Since the two daughter subtrees are completely independent, this probability is
given by their product.

Note that the recursion relation in Eq. (\ref{drecur1}) for the height distribution 
$Q_l(n)$ in the b-DST is identical to the recursion in Eq. (\ref{recur1}) in Section II for the
probabilities $G_l(n)$ in the b-DDLA problem, except for the slight difference in the
initial values $Q_1(n)$ and $G_1(n)$. This slight difference does not affect the
asymptotic behaviors. So, one can apply all the results obtained via the traveling front approach
in Section II for the b-DDLA model directly to the b-DST problem. In particular, the result
in Eq. (\ref{l*2}) indicates that the average height of the b-DST has the asymptotic 
following behavior for large $n$,
\begin{equation}
l^*(n) \approx {\log}_2(n)+ \sqrt{{2\over {b}}{\log}_2(n)}.
\label{dl*2}
\end{equation}
For $b=1$, this result coincides with that of Aldous and Shields\cite{AS} obtained
by probabilistic methods. Note that for $b=1$ case, a more refined result including 
additional subleading terms to Eq. (\ref{dl*2}) was recently obtained in Ref.\cite{KS}
using rigorous methods. However, for general $b$, we are not aware of any rigorous
results in the computer science literature and our Eq. (\ref{dl*2}) seems to be the
first result for the average height of a b-DST. Furthermore, Eq. (\ref{wn}) in Section II
predicts that the standard deviation of the height around its average value decays 
extremely slowly with large $n$, $w(n)\approx 1/{\sqrt{2b\ln (2)\ln (n)}}$. This result
on the variance of the height in b-DST also seems not to have been obtained by other
methods before.   

We now turn to the Lempel-Ziv algorithm for data compression\cite{LZ}. The connection between 
this algorithm and the DST problem was known before\cite{AS,LS}. The Lempel-Ziv algorithm 
is central to many universal data compression schemes and have many applications such as  
in the efficient transfer of data\cite{LS}. This basic scheme of 
this algorithm is very simple:
it takes a given data string, say a sequence of binary digits such as $11000110111011110$,
and partitions it into `words'. `Words' are  subsequences of variable sizes which are 
never repeated and are constructed by employing the rule that a new `word' is the shortest
subsequence not seen in the past as a `word'. This is best understood by an example.
\begin{figure}
  \narrowtext\centerline{\epsfxsize\columnwidth \epsfbox{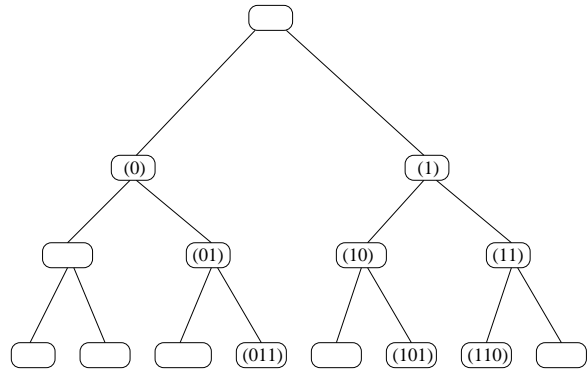}}
\caption{The figure shows how the partitioning of a sequence
$11000110111011110$ into `words' $(1)(10)(0)
(01)(101)(11)(011)(110)$ using the Lempel-Ziv parsing algorithm can be represented
as a digital search tree. The length of a `word' is equal to its depth in the tree
measured from the empty root at the top.}
\end{figure}
\noindent
Consider the binary sequence $11000110111011110$ and construct `words' starting from the
left end using the Lempel-Ziv algorithm. Starting from the left end, the first digit encountered
is $1$. Since $1$ has not occurred before as a `word', one can form the first `word' $(1)$. Now 
we move to the next element which also happens to be $1$. But, now since $(1)$ is already a `word', 
the shortest segment we can use to form a `word' is $(10)$. Similarly, the next word would be 
$(0)$ since $(0)$ has not occurrred before as a `word'. One keeps repeating the procedure and at 
the end, the original sequence is partitioned into the following sequence of `words': $(1)(10)(0)
(01)(101)(11)(011)(110)$. The original data is thus compressed into these words. Even though the 
`words' are relatively short in the begining, it turns out that they become bigger quite
rapidly. One of the interesting questions of practical importance in this scheme
is the statistics of the longest `word' when the original data string is random. For 
concreteness,
let us consider a random binary sequence of initial data and focus on the first
$n$ `words'. Let $l$ be the length of the longest `word' amongst these $n$ words.
Clearly $l$ is a random variable since the underlying binary sequence is random.
We are interested in the statistics of $l$ as a function of $n$.

There is a natural representation of this parsing algorithm in terms of a DST\cite{LS}.
Consider a binary tree whose nodes are initially empty. In fact, the root of this
tree is always going to be empty (see Fig. 4). 
Now we take the first of the Lempel-Ziv
parsed `words' and examine its first digit. If the first digit of this `word' is $1$,
we put this `word' in the right daughter node of the root. On the other hand, if the first digit
is $0$, we put this word at the left daughter node. This newly occupied node is now full
and can not accomodate any other `word'. Then we consider the second `word' and look at its first 
digit. If the first digit is $1$ ($0$), we go to the right (left) daughter node. 
Let us call this node $i$. If this
node $i$ is empty, we put the `word' there. If $i$ is already occupied by the first `word', then
we need to examine the second digit of our second `word' and depending on its value ($1$ or $0$),
we go respectively to the right or to to the left daughter node of $i$ and put our second `word'
at this new site. 
This process is repeated until all the words are stored and the resulting
tree is clearly a DST, since at each step the decision to go to the left or to the right occurs
randomly (due to the randomness of the underlying binary sequence where each digit can
be either $0$ or $1$ with equal probability). The construction of this DST from the
parsed words $(1)(10)(0)(01)(101)(11)(011)(110)$ is shown in Fig. 4.         

It is clear from the algorithm that the depth of a given `word' in the DST (measured from the 
empty root) is precisely the equal to the length of the `word' (see Fig. 4). In particular,
the longest `word' will also be the farthest from the root. Thus the length $l$ of the longest 
word is precisely the height of the corresponding DST. There is a generalized Lempel-Ziv
algorithm where during the partitioning into `words', any particular `word' is allowed to be 
repeated at most $b$ times\cite{LS}. Then the corresponding DST is precisely a b-DST. Thus our 
results regarding the average height $l^*(n)$ and its width apply as well to the
longest `word' in the generalized Lempel-Ziv algorithm.  

\section{Conclusions}
In this paper we have used a suitably adapted version of the traveling front approach
to derive exact asymptotic results for the statistics of the number of particles 
in a generalized directed diffusion limited aggregation problem. We have pointed
out a close connection of this problem to two separate problems in computer science, namely
the digital search tree problem and the Lempel-Ziv algorithm used for data compression.
Our results for the number of particles in the generalized b-DDLA model have direct
relevance to the statistics of height in the digital search tree problem and to
the statistics of the longest word in the Lempel-Ziv algorithm. 

The traveling front approach has recently been used successfully\cite{bst1,bst2} to derive 
exact asymptotic results for heights in a number of growing search tree problems in
computer science. This paper shows that the scope of this approach can be extended 
to include yet another different class of search trees, namely the digital search tree.
The main advantage of this method is that it provides an easy way to derive the leading 
asymptotic behavior exactly in a variety of extreme value problems\cite{TF}.

The present study leads to a number of interesting, open problems which we list below.

{\em Undirected DLA problem on a tree}: In this paper, we have focused on 
a directed model for simplicity, where the particles undergo diffusion but only in the overall 
downward direction. It would be interesting to extend the traveling front approach to
an undirected model such as the one studied by Hastings and Halsey\cite{HH}.

{\em Subleading Corrections in the biased DDLA problem}: Using traveling front approach, we 
managed to calculate only the leading behavior of the average density in the biased DDLA
model (see Section IV). It would be interesting to compute the subleading corrections
to this leading behavior.

{\em DDLA model with stochastic screening}: In this paper, we have studied a DDLA
model where the screening is deterministic in the sense that
a particle, on its way downwards, stops definitely when it reaches at a site such that at least
one of the daughter nodes of that site is occupied. It would be interesting to consider a stochastic
screening version defined as follows. For simplicity, we define the model for $b=1$ case, though 
it can be trivially generalized to $b>1$ case also. The particles are introduced sequentially
from the top as before and a new particle is introduced only when the previous particle has
completely
stopped moving. On its way down, at each site $i$ a particle performs the following steps:
if both the daughter nodes of $i$ are already occupied, the particle rests at $i$ for all
subsequent times and then a new particle is added. If both the daughter nodes are empty, then
the particle chooses one of the daughter nodes at random and moves there. If, however, only one
of the daughter nodes (say the left one) is occupied but the other one (the right one) is
empty, then with probability $p$ ($0\leq p\leq 1$) the particle moves to the right daughter node
and with probability $(1-p)$ it rests at $i$ for all subsequent times.
Clearly the case $p=0$ corresponds to the deterministic screening model studied in Section II.
On the other hand, for $p=1$, it is obvious that at the saturation the tree will be
completely full and the density will be exactly $1$.
It would be interesting to compute the statistics of density in this stochastic screening model.

{\em Disordered b-DDLA model}: In the present paper, we considered the b-DDLA model
when all the nodes of the tree have the same capacity $b$. In a disordered version
of the problem, this node capacity $b_i$ of a site $i$ may vary from one site to 
another. One can consider
$b_i$'s to be a set of quenched variables (as in the usual models in disorderd systems),
each drawn independently from a specified distribution $p(b)$. Then, for a given fixed set 
of 
${b_i}$'s, one would first like to compute, for example, the average number of particles 
$\langle n_l\rangle$ at saturation, and then average this quantity over the disorder
to obtain ${\overline {\langle n_l\rangle} }$, where ${\overline {...}}$ denotes the average over 
the ${b_i}$'s.  It 
would also be interesting to compute
the sample to sample fluctuations of the average density $\langle n_l\rangle$.

\section*{Acknowledgements}

I thank A.J. Bray, D.S. Dean, D. Dhar and P.L. 
Krapivsky for discussions. I thank P.L. Krapivsky also for pointing out the reference 
\cite{BS}. The hospitality of the Tata Institute, Bombay, where part of this work was done
during a visit, is gratefully acknowledged.

\appendix 
\section*{Derivation of the Generating Function}

In this appendix, we present the derivation of Eq. (\ref{fls1}) where ${\tilde F}_l(s)$
is defined in Eq. (\ref{gf1}).
Our starting point is the linear equation (\ref{recur3}). We first define the exponential 
generating function,
\begin{equation}
H_l(z)=\sum_{n=0}^{\infty} F_l(n) {z^n\over {n!}}.
\label{hlz}
\end{equation}
Multiplying both sides of Eq. (\ref{recur3}) by $z^n/n!$ and summing over $n$, it is easy to see
that $H_l(z)$ satisfies the $b$-th order nonlocal differential equation,
\begin{equation}
{{d^b H_{l+1}(z)}\over {dz^b}}= 2 H_{l}(z/2)e^{z/2},
\label{dhlz}
\end{equation}
for all $l\geq 1$. This recursion in Eq. (\ref{dhlz}) starts from the initial function
$H_1(z)$ which needs to be computed separately. Using $F_1(n)=0$ for $0\leq n\leq (b-1)$ 
and $F_1(n)=1$ for $n\geq b$, we find $H_1(z)= \sum_{k=0}^{b-1}z^k/k!$. The next step
is to define a new function, 
\begin{equation}
U_l(z)= H_l(z)e^{-z}=\sum_{n=0}^{\infty} F_l(n) {{z^n}\over {n!}}e^{-z}. 
\label{ulz}
\end{equation}
From Eq. (\ref{hlz}), 
it follows, after a few steps of algebra, that $U_l(z)$ satisfies the differential equation,
\begin{equation}
\sum_{k=0}^b {b\choose k} {{d^k U_{l+1}(z)}\over {dz^k}}= 2 U_l(z/2),
\label{dulz}
\end{equation}
for all $l\geq 1$ starting with the initial function, $U_1(z)= e^{-z}\sum_{k=0}^{b-1}z^k/k!$.

We now define the Laplace transform, ${\tilde U}_l(s)=\int_0^{\infty} U_l(z) e^{-sz} dz$.
Taking the Laplace transform in Eq. (\ref{ulz}), we get
\begin{equation}
{\tilde U}_l(s)= \sum_{n=0}^{\infty} F_l(n) {1\over {(1+s)^{n+1}}}={\tilde F}_l(s),
\label{uls}
\end{equation}
where we have used the identity $\int_0^{\infty} e^{-z}z^{n}dz =n!$ and the definition
of ${\tilde F}_l(s)$ in Eq. (\ref{fls1}). Next we take the Laplace transform on both
sides of Eq. (\ref{dulz}). Using the initial conditions for $n=0$, one can show easily that
$d^k U_l(z)/dz^k |_{z=0}=0$ for all $l\geq 1$ and $k\leq (b-1)$. Using this condition and doing
integration by parts, one finds 
\begin{equation}
\sum_{k=0}^b {b\choose k} s^k {\tilde U}_{l+1}(s)= 4\, {\tilde U}_{l}(2s).
\label{uls1}
\end{equation}
Summing the left hand side of Eq. (\ref{uls1}) and identifying ${\tilde U}_l(s)={\tilde F}_l(s)$
as in Eq. (\ref{uls}) then gives the desired recursion relation
\begin{equation}
{\tilde F}_{l+1}(s)= {4 \over {(1+s)^b}} {\tilde F}_{l}(2s).
\label{desired}
\end{equation}

\end{multicols}
\end{document}